# Entropy and spontaneity in an introductory physics course for life science students

Benjamin D. Geller, Benjamin W. Dreyfus, Julia Gouvea,
Vashti Sawtelle, Chandra Turpen, and Edward F. Redish

*Department of Physics, University of Maryland, College Park, MD 20742*

**Abstract.** In an Introductory Physics for Life Science (IPLS) course that leverages authentic biological examples, student ideas about entropy as "disorder" or "chaos" come into contact with their ideas about the spontaneous formation of organized biological structure. It is possible to reconcile the "natural tendency to disorder" with the organized clustering of macromolecules, but doing so in a way that will be meaningful to students requires that we take seriously the ideas about entropy and spontaneity that students bring to IPLS courses from their prior experiences in biology and chemistry. We draw on case study interviews to argue that an approach that emphasizes the interplay of energy and entropy in determining spontaneity (one that involves a central role for free energy) is one that draws on students' resources from biology and chemistry in particularly effective ways. We see the positioning of entropic arguments alongside energetic arguments in the determination of spontaneity as an important step toward making our life science students' biology, chemistry, and physics experiences more coherent.

## I. INTRODUCTION

### Why Entropy?

Countless physical processes that *could* spontaneously occur without violating the conservation of energy principle are never observed. We do not see chairs spontaneously absorb energy from the floor and begin to slide across rooms. We do not see smoke spontaneously coalesce in the corners of smoky bars. And we do not see shivering campers transfer heat to their campfires. None of these processes would violate the first law of thermodynamics, but all of them violate the second. An introductory physics course that emphasizes the first law of thermodynamics but gives short treatment to the second may help students make sense of why only energy-conserving thermal processes are ever observed, but it does not provide students with an opportunity to make sense of why so many more energy-conserving processes are not. Put another way, the first law of thermodynamics is a necessary but insufficient rule for making sense of the thermal world.

There is another motivation for including a robust discussion of entropy and the second law in introductory physics classrooms – the second law makes contact with our everyday intuitions about energy in a way that the first law does not. In modern sociopolitical discussions, energy is "wasted" and "used up," and we worry a great deal about how to "conserve it."[1] Since the first law of thermodynamics guarantees conservation of energy, students in a course that emphasizes only that law might well perceive a disconnect between energy as discussed in physics classrooms and energy as discussed in their everyday lives.[2] Effective teaching demands that we leverage students' everyday ideas about the world around them, and the second law is uniquely positioned to help unpack the everyday idea that not all energy is equally useful. Although the amount of total energy in the world does not diminish, the amount of *useful* energy does.[3] Only the Second Law of Thermodynamics accounts for this essential distinction

### Why entropy in an IPLS course?

The case for including entropy in any introductory physics treatment of thermodynamics is strong, but the case is even stronger for including such a treatment in Introductory Physics for Life Science (IPLS) courses. Randomness and diffusive processes are of particular importance in biological systems, and a deep understanding of such processes requires a facility with the Second Law. Diffusion accounts for the movement of oxygen from the alveoli to the capillaries during respiration, and for the movement of $CO_2$ within leaves for use in photosynthesis.[4,5] The formation of ordered biological structures in aqueous environments depends on an entropically driven hydrophobic effect.[6] And an understanding of how directed motion can emerge from random motion is essential for making sense of many directed sub-cellular processes.[7]

A second reason for spending considerable time on entropy in an IPLS course is that it provides the link between energy as described in a typical physics course and free energy as described in a typical biolo-gy or chemistry course.[8] Consider the relationship between the Helmholtz free energy, the internal energy, and the entropy of a system:

$$F = U - TS$$

In biology and chemistry texts, it is often the free energy *F* (or, as we will see later, the Gibbs free energy *G*) that plays a central role, as the sign of the change in that quantity determines whether biochemical processes and reactions can occur spontaneously.[8] In a typical introductory physics treatment of energy, some time is spent unpacking what goes into the internal energy term *U* in the above equation. A treatment of electrostatics, in particular, can be viewed through a thermodynamic lens as living inside the internal energy term in the expression for free energy. (Unfortunately, in a typical introductory course, "conservation of energy" and "the First Law of Thermodynamics" are usually taught separately, and electrostatics is never explicitly connected to thermodynamics. Seeing electrostatics as living inside *U* requires a layer of interpretation not often apparent to our students.) The link between energy and free energy is entropy, and as the entropy of a system increases, less energy is "free" to do useful work.[8]

To make the connections between energy, entropy, and free energy concrete, consider a standard idealized physics problem, the free expansion of a thermally isolated ideal gas (Figure 1).

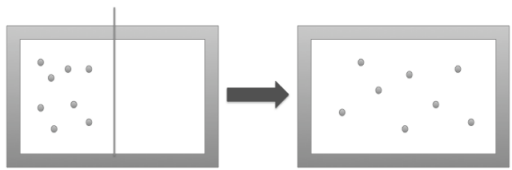

**Figure 1**. Free expansion of a thermally isolated ideal gas. When the barrier is removed, the gas expands to fill the available volume. The *energy* of the gas is constant during this isothermal free expansion, but the *free energy* of the gas decreases.

When the barrier is removed, the gas freely expands to fill the available volume. Since the compartment is thermally isolated from its surroundings, the internal energy *U* of the gas remains constant during the expansion, while the entropy of the gas increases. The result is that the freely expanded gas has the same energy but less *free* energy than the gas did before the barrier was removed. The expanded gas has less capacity to do work on its surroundings than the compressed gas did, which makes sense if we make the typical association between free energy and the capacity to perform useful work.[8]

This paper argues that thinking of entropy as a link between energy and free energy, and in turn framing a discussion of the Second Law of Thermodynamics in the context of considerations about free energy and spontaneity, can be an important step toward bridging different disciplinary treatments of thermodynamics. The context for this discussion is the NEXUS/Physics course,[9,10] an introductory course for life science students that leverages students' experiences in introductory biology and chemistry courses. Meeting our students where they are means building upon the resources for thinking about free energy and spontaneity that IPLS students bring from their experiences in those classes. Many students enrolled in NEXUS/Physics had not previously taken a chemistry course with an *explicit* focus on entropy and free energy. Despite this, our students report having seen these ideas in their introductory biology and chemistry courses, and report having used these ideas in meaningful ways. In light of these reports, we see the discussion in this paper as relevant to a wide range of IPLS courses, including ones in which students may not yet have had courses that explore thermodynamics in depth.

## II. METHODOLOGY

We draw on case study interviews conducted with students in our NEXUS/Physics course at the University of Maryland, College Park (UMCP).[9,10] Because our course has a year of biology and a semester of chemistry as prerequisites, our assignments and small group problem solving sessions leverage our students' familiarity with the material in those courses by introducing authentic biological problems from the beginning. The second iteration of NEXUS/Physics was offered to 31 students during the 2012-2013 academic year.

In order to get a sense for how our students had previously been exposed to entropy, free energy, and spontaneity, we interviewed six students prior to the unit about entropy and the second law of thermodynamics, and again once or twice (depending on the student) after the unit was complete. Our focus in this paper is on four of these case study students: Elena, Tammy, Gavin, and Otto (all pseudonyms). We chose to focus on these four students because all of them demonstrated nuanced conceptual resources for thinking about entropy and its relation to free energy and spontaneity. We focus in particular on interviews with these students *prior* to the unit on entropy, at which point their views had likely been shaped primarily by their experiences in either high school or in biology and chemistry courses at the University of Maryland.

Although our interviews with these students suggested that their thinking about entropy and related topics had changed as a result of the tasks and problems completed during our Second Law unit, we do not examine those changes in any detail here. We also make no quantitative claims about the generality of these four students' responses. Instead, we describe how these students' notions about entropy and spontaneity from prior experiences suggest an approach to teaching these topics in an IPLS environment. Our aim is to describe how the ideas expressed by Elena, Tammy, Gavin, and Otto may inform our efforts to make life science students' physics, biology, and chemistry experiences more coherent.

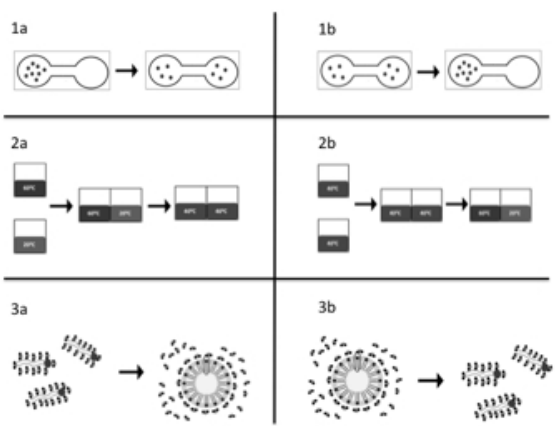

**Figure 2**. Images of slides shown to students during interviews prior to the entropy unit in NEXUS/Physics. Box 1a shows particles filling the available volume during the free expansion of a thermally isolated ideal gas. Box 2a shows two objects initially at different temperatures brought into contact, and ultimately equilibrating to a common temperature. Box 3a shows phospholipid molecules, initially each surrounded by water molecules, self-assembling into an organized micelle. Boxes 1b, 2b, and 3b show the same three processes happening in reverse.

To get a sense for how our case study students thought about entropy, we showed each of them a series of slides illustrating processes for which entropy plays a significant role (Figure 2).

We showed the students three processes – the free expansion of an ideal gas, the equilibration of two objects at different temperatures, and the formation of micelles out of phospholipids in water – and asked the students to describe what they observed happening in each slide. We also showed the students the same three processes in reverse, hoping to prompt discussion about why the forward direction was spontaneous but the reverse was not.

In the case of micelle formation, the water molecules have more degrees of freedom when not bound to individual phospholipids, contributing to an overall increase in the system's entropy when the micelle forms and lipid-water interactions are minimized. Because the net energetic change associated with bond breaking and reforming events in micelle formation is nearly zero, this entropic effect drives the process.[6]

To provide some context for our case study students' ideas about entropy, and to get a feeling for how entropy was being discussed in other disciplines, we examined the treatment of entropy in the introductory biology, chemistry, and physics textbooks used by life science students at the University of Maryland, College Park (UMCP).[11,12,13] Such disciplinary textbooks differ in both the language they use to describe entropy, and in the examples used to illustrate why entropy is important. The discussion here is in no way meant to be an exhaustive account of the sorts of treatments of entropy and the Second Law that one might find in various disciplines' textbooks, but rather serves to present typical examples of each discipline's treatment in the hopes of providing context for making sense of our students' ideas.

### III. THERE ARE TENSIONS IN STUDENTS' REASONING ABOUT ENTROPY AND SPONTANEITY

In analyzing student interviews, our focus was on understanding the ideas and resources our students had for thinking about entropy and for using the Second Law to describe phenomena in the natural world. Prior work has described a number of patterns in undergraduate students' thinking about both entropy[14-16] and spontaneity,[17-18] but few attempts have been made to understand student thinking about the Second Law in biological contexts.[19] We uncovered tensions between what Tammy, Otto, Elena, and Gavin understood to be true about entropy and what they knew to be true about the spontaneous formation of biological structures. At times students recognized these tensions in their own understanding, and at other times we identified tensions that students did not fully articulate themselves.

#### Students experience tension between disorder and biological structure

Prior to beginning any discussion of entropy in our NEXUS/Physics course, we asked our case study students to tell us how they had seen the concept described in their previous biology and chemistry coursework. Their initial responses were fairly uniform, all leveraging ideas about disorder and chaos, and entirely consistent with the descriptions found in standard introductory textbooks.

Gavin said that "entropy is a measure of disorder," and that "generally, the universe is increasing in disorder." Tammy and Otto described entropy as related to "chaos," with Otto referring to an entropy increase as "the increasing chaos or disorganization [in the world]," and Tammy defining entropy as "the amount of chaos that is related to the system." Tammy further specified what she meant by chaos, saying that "[chaos] can be determined by things like the number of molecules, whether a structure is forming or 'disforming'... the type [phase] of matter: solid, liquid, or gas…" Elena was at first uncertain about whether to associate entropy with order or disorder, but was confident that somehow entropy referred to the "order and disorder of a system."

Otto later revised his initial definition of entropy to include what he remembered to be true about the entropy change associated with chemical reactions, saying that, in addition to relating to chaos, entropy is also associated with "breakdown," as when a single mole of some substance breaks into two moles. By way of illustrating his point, Otto wrote down a chemical reaction – the decomposition of carbonic acid – and noted that "there is more entropy [on the right side of the reaction] because you go from one mole of a compound to two moles... it is broken down... so that increases the entropy [just because you have more moles]." Otto even found ways of blending these "chaos" and "breakdown" metaphors for entropy, describing both the chaos of diffusion and the breakdown of chemical compounds as characteristic of processes in which one "loses containment" of the system.

These initial responses were not surprising. All of the case study students recalled having encountered entropy in a manner consistent with introductory textbook treatments of the topic. In the introductory biology textbook used by our students, for example, entropy is defined simply as "a measure of disorder," with no more precise description or quantitative representation provided.[11] The book notes only that "the more randomly arranged a collection of matter is, the greater its entropy," and that "there is an unstoppable trend toward randomization of the universe as a whole." The introductory chemistry textbook assigned to our students in their general chemistry course introduces entropy in a similar way as the biology textbook, again defining entropy as a measure of disorder in a substance or system.[12] One notable way in which the chemistry text differs from the biology text is in its inclusion of a table of absolute molar entropies for various substances, a table whose values are consistent with Otto's understanding that entropy increased upon the breakdown of carbonic acid. When entropy is described qualitatively in the introductory physics text used by students at UMCP who are not enrolled in NEXUS/Physics, it also does so with the familiar refrain that "entropy measures the amount of disorder in a system."[13]

For some students, this idea of entropy as disorder is in tension with their knowledge that organized biological structures spontaneously form. Consider, for example, how Gavin talks about his understanding of micelle formation, the process by which phospholipid molecules self-assemble into an organized spherical structure in which the polar heads of the molecules interact with water and the non-polar tails are buried inside (see Boxes 3a and 3b in Figure 2):

> *We say that it is thermodynamically favorable for entropy to increase… then why is it that you have situations where cells are going to congregate? [Where] you are going to make organisms? [Where] you are going to make people?... What I know about entropy is what I have been taught… I do not have all the information yet. I have been taught that [micelle formation] happens, I just don't know how it happens.... I know the fundamental properties of these molecules and how they interact with each other but… if you are going to want everything to spread out, then you are going to assume that everything will spread out... but you know the complete opposite happens where it becomes more organized… I feel it disagrees. I feel like it is a lack of complete information. [I have] enough to answer a question on the MCAT but not enough information to have a symposium about how micelles form.*

Accounting for such organized biological structures against the backdrop of the ubiquitous "entropy as disorder" metaphor requires that we think carefully about how to connect ideas about entropy to ideas about spontaneity. Before we turn to this issue and how we might address it, however, it is worth emphasizing that *any* such treatment will have to take seriously this tension that Gavin articulates. It is natural that IPLS students would feel tension between the second law of thermodynamics and what they know to be true about spontaneous cellular structure formation. While it may be technically sufficient to address the tension by asking students to think about the entropy of the universe as a whole rather than just the entropy of the system, it is an empirical question as to whether such an approach in and of itself best leverages student resources for thinking about spontaneity. As we will see, coordinating such an approach with one that addresses the relationship between system free energy and spontaneity may in fact make use of student resources in a promising way.

## Tensions exist between colloquial and technical meanings of disorder

There is a second tension that we must consider alongside the tension between disorder and organized biological structure, one that is embedded within the disorder metaphor itself. While the terms "disorder" and "chaos" may well have a very technical meaning to scientists teaching the second law of thermodynam-ics, people generally mean a wide variety of less precise things when using those words.[20-24] To see how the technical meaning of "disorder" in the context of the second law of thermodynamics need not align with other plausible meanings of the word, consider how Elena described the PowerPoint slide showing a gas freely expanding to fill a volume (Box 1a in Figure 2). She considered the final state to be more "ordered" because she associated order with "the natural state of things." She identified the freely expanded gas as "a more ordered system because [the expansion] would happen naturally... it does not go against nature." While Elena's definition of "order" may contradict the technical sense in which the word is sometimes intended, there is actually nothing linguistically implausible about associating "order" with the natural state of things. Indeed, Elena's mechanistic reasoning about the free expansion is entirely sound: "All the molecules are not going to want to stay in one space, because they are going to be interacting with each other and bumping into each other. [Individually] they will be moving to the right and to the left... and eventually they are going to equilibrate and that [spread out state] is just more ordered."

Elena's description of the free expansion of an ideal gas reveals one pitfall of using a term that has numerous colloquial meanings. Given the imprecise definition of "disorder" and the variety of meanings students may associate with the word, there is no obvious reason to expect students to associate "disorder" with one particular meaning and not others.

Tammy's interview further illustrates this point in her description of the freely expanding gas. Consistent with her definition of chaos as involving more violent and numerous collisions, Tammy said that the gas "has more entropy [before it expands] because... you have more molecules in a smaller volume… and therefore more collisions… If you think about 8 molecules colliding in a tiny container versus 8 molecules colliding in a giant container, the giant container is not going to have as much going on because the molecules are so far [apart], whereas in the smaller container it's much more chaotic... much more going on." In short, Tammy says that the gas freely expands because "it is moving from a more chaotic to less chaotic system, which is more favorable… more balanced."

Tammy's perfectly reasonable sense of what it means for a system to exhibit chaos (indeed it is a meaning that instructors would likely promote in other scientific contexts), does not match the narrow technical sense in which instructors mean "chaos" when describing entropy. As a result, Tammy sees the freely expanding gas as becoming less chaotic and must conclude that the entropy decreases in such a process.

One way to resolve the tension between technical and colloquial uses of "disorder" and "chaos" would be to stop using those metaphors entirely when discussing the Second Law of Thermodynamics. This approach, however, not only disregards the entrenchment of those terms in everyday understanding of entropy, but it ignores the highly productive ways in which students leverage the terms. In the next sections we highlight some of these productive uses of disorder, and suggest that refining the metaphor and coordinating it with other formulations of entropy may be more effective than trying to eliminate it from our vernacular entirely.

## IV. STUDENTS HAVE PRODUCTIVE RESOURCES FOR MAKING SENSE OF ENTROPY AND SPONTANEITY

Meeting our life science students where they are means leveraging the resources that they bring to IPLS courses from their experiences in biology and chemistry courses. In this section we describe some of the resources that our students have for making sense of entropically driven processes, and for thinking about the relationship between entropy and spontaneity. We find that the *free energy* of a system – and in particular the way in which energetic and entropic effects determine the change in free energy of a system – plays a central role in our students' understanding of thermodynamics.

### Using disorder to explain diffusion and osmosis

In describing what he meant by chaos, Otto appealed to the process of osmosis. He noted that, as solvent water molecules spread toward regions of high solute concentration, the "water is no longer contained... it is distributed... it is all over the place... you can think of it as a mess." Later he described the water in such a process as going from "a place of order to a place of disorder." For Gavin, the idea of disorder is central to his understanding of diffusion: "[If you] put a bunch of particles together in a certain area, the particles want to diffuse from one another; they want to spread out; they do not want to be so ordered; they want to increase [their] disorder."

To Gavin and Otto, processes like diffusion and osmosis make good sense in the context of metaphors like "disorder" or "chaos." The metaphors are doing productive work for them in making sense of entropic phenomena. This would suggest that for Tammy and Elena, whose plausible ideas about "disorder" did not align with the technical meaning of the word in the context of the Second Law, the challenge is not to replace the word but to refine its definition so as to disambiguate colloquial from technical meanings, and to help them understand the conditions under which certain colloquial meanings are appropriate. A nuanced and technically sound understanding of disorder (perhaps one that is well coordinated with canonical ideas about microstates, for example) would still, however, fail to address the tension between disorder and the spontaneous appearance of order in the biological world. To address that tension, student ideas about the interplay of energy and entropy are particularly relevant.

### Spontaneity depends on both energy and entropy

In one way or another, all of the students we interviewed related their understanding of entropy to their understanding of energetic interactions between molecules and, in turn, to Gibbs free energy. When Gavin was confronted with the apparent conflict between his idea of entropy as disorder and his knowledge that micelles form spontaneously in water, Gavin's first looked for an energetic argument that might help with the reconciliation. "If you threw 100 polar molecules in the ocean," Gavin said, "then over time they would spread out as far as they can possibly get until they aren't considered interacting with each other any more." He attributed the observation that such separation does not occur for non-polar lipid molecules to the fact that "entropy [must be considered] relative to interactions... it is dependent on how much force [the molecules] can influence on each other." He then spontaneously brought up the equation relating entropy to Gibbs free energy,

$$\Delta G = \Delta H - T\Delta S,$$

and noted that "the higher the entropy, the more negative the free energy, depending on enthalpy in the system… the more spontaneous something is, the higher the entropy in the system." Gavin looked to an interplay between entropic and energetic effects to help him reconcile micelle formation with what he would expect from entropy considerations alone.

Likewise, Otto makes sense of the apparent contradiction between "entropy as chaos" and the spontaneous formation of a micelle in water by noting that "naturally things just want to go into chaos... but [in micelle formation] that is not really the case because of the polarity that is involved. It is the polarity that is causing it to come together compared to just having things naturally interact with each other." Otto has not carefully thought through the energetic factors involved in the interactions between lipids and water molecules, but his intuition, like Gavin's, is that one must consider those energetic factors alongside the entropic effects in determining whether a structure will spontaneously form.

### Using $\Delta G$ and spontaneity as a check on ideas about disorder

Tammy, who had a common sense definition of "disorder" that gave entropy the wrong sign, also quickly turns to the $\Delta G = \Delta H - T\Delta S$ relation, using it as a check on her previous conclusion that the entropy change would be negative upon free expansion of an ideal gas. She recognizes that a negative $\Delta S$ corresponds to a positive $\Delta G$ in situations where the enthalpy change is zero, but she knows that the free expansion requires a negative $\Delta G$. "It bothers me," Tammy says, "because [free expansion] is a spontaneous process and [a positive $\Delta G$] means it is not spontaneous... it is saying that it requires energy… something is overlooked... whether that means there's actually an enthalpy change I am not sure. I am really sure that the $\Delta G$ should be negative... because, you know, it is a spontaneous process." Later in the interview, Tammy's confidence in her reasoning about free energy forces her to conclude that she must have been reasoning incorrectly about the sign of the entropy change.

For Tammy and all the other students we interviewed, the idea that spontaneity requires a negative change in the Gibbs free energy of the system served as a powerful resource for framing a discussion about what entropy can contribute to their understanding of biological phenomena. In fact, across all of our interviews with the case study students, the notion that spontaneity requires a negative change in the Gibbs free energy of a system was one of the most consistently leveraged ideas, and was well coordinated with other elements of our students' thermodynamic knowledge.

## PATHWAYS TOWARD BRIDGING ENTROPY AND SPONTANEITY

The life science students we interviewed had powerful resources for reasoning about spontaneity. The goal in this section is to describe how these ideas can be positioned relative to canonical statements of the Second Law, and to point toward ways in which one might leverage these ideas in scaffolding tasks that support developing students' understanding. Our goal is *not* to offer a one-size-fits-all approach to addressing the tensions described earlier in this paper. Rather, we discuss how our students' familiarity with free energy suggests one promising route toward bridging ideas about entropy and spontaneity, one that foregrounds a statement of the Second Law in terms of energetic and entropic changes in the *system*.

## Two ways of thinking about the Second Law of Thermodynamics

The Second Law says that a physical process is spontaneous if it is associated with positive change in the overall entropy of the *universe*. Use of this formulation of the Second Law to predict spontaneity is limited, however, by ones ability to account for all the entropy changes in the universe during a given process. Fortunately, under certain conditions one can re-write the Second Law such that spontaneity is determined by a property of the *system*, and not by a property of the universe as a whole. At constant temperature ($T$) and pressure ($P$), conditions common for biochemical processes, the system property that determines spontaneity is the Gibbs free energy. The Gibbs free energy differs from the Helmholtz free energy in that the former is a measure of the amount of work that one can obtain from a thermodynamic system at constant $T$ and $P$, whereas the latter measures the obtainable work when only $T$ and $V$ are constant.[25] Figure 3 demonstrates the relationship between the entropy of the universe and the Gibbs free energy of a system.

When the entropy of the universe increases during a process, the Gibbs free energy of the system decreases, and the process is spontaneous. When the entropy of the universe decreases during a process, the Gibbs free energy of the system increases, and the process does not spontaneously proceed.

The relationship between the system's Gibbs free energy change and the universe's entropy change suggests two possible ways of connecting ideas about disorder with the spontaneous formation of organized structure. On the one hand, biological structure formation can be reconciled with the Second Law of Thermodynamics by considering not just the entropy of the system, but also the entropy of the surroundings (line 2 in Figure 3). This approach requires that one have some way of measuring entropic changes not just in the local system one is investigating, but *everywhere*. In predicting the spontaneity of micelle formation, for instance, one would not only need to measure the entropic changes for the phospholipid and water molecules in the system, but also for everything else in the universe that was changed by the process being considered. This is an approach to conceptual reconciliation that has been previously discussed in the literature,[26] and one that many physicists are familiar with employing when confronted with the question of how organized structures form.

A second approach to reconciling biological structure formation with the Second Law of Thermodynamics is to consider the interplay of energetic and entropic effects on the *system itself* (line 1 in Figure 1). Our case study interviews with students prior to the entropy unit in our course suggest that this approach may draw on students' experiences in biology and chemistry in particularly effective ways. Indeed, the introductory biology and chemistry textbooks that our students use *introduce* entropy by way of its contribution to the Gibbs free energy.[11,12] The very first time that the symbol $S$ appears in their biology text is in the equation $\Delta G = \Delta H - T\Delta S$, wherein an increase in entropy is one way to achieve a decrease in Gibbs free energy $G$.[11] The introductory chemistry textbook introduces entropy in a very similar way, positioning entropy as one of the factors that one must consider in determining the sign of $\Delta G$.[12]

$$\text{For processes at constant } T \text{ and } P:$$
$$\Delta G_{sys} \equiv \Delta H_{sys} - T\Delta S_{sys} \quad \text{(line 1)}$$
$$= -T\Delta S_{surr} - T\Delta S_{sys} \quad \text{(line 2)}$$
$$= -T\Delta S_{univ}$$

A positive value for $\Delta S_{univ}$ corresponds to a negative value for $\Delta G_{sys}$

**Figure 3**. For any process that occurs at constant temperature ($T$) and pressure ($P$), the Gibbs free energy of a system changes in a way that mirrors the entropy change of the universe. As such, the sign of the system's Gibbs free energy change during a process determines whether the process is spontaneous. Line 2 is equivalent to line 1 because the enthalpy change for a system at constant $T$ and $P$, $\Delta H_{sys}$, is equivalent to the heat transferred with the surroundings, $-T\Delta S_{surr}$. Ref. 25 provides more detailed relations between enthalpy, heat transfer, and $T\Delta S$.

By way of contrast, the introductory physics textbook used by life science students not enrolled in NEXUS/Physics never mentions the role that entropy plays in determining the change in free energy during a physical process. In fact, despite its central role in the biological and chemical sciences, free energy is not considered anywhere in the entire introductory physics textbook. By describing the Second Law not just in terms of the entropy of the universe, but also in terms of the Gibbs free energy in a system, IPLS courses can therefore play a critical role in bridging the divide between canonical treatments of entropy in different disciplines.

## Leveraging students' ideas: A sample activity

The student data in our IPLS course suggests that meeting our students where they are means leveraging their familiarity with Gibbs free energy in making sense of the Second Law of Thermodynamics. Fortunately, a treatment of spontaneity that emphasizes the interplay between energy and entropy is one for which an introductory physics course is naturally well-suited. Unpacking the mechanistic underpinnings of entropy and enthalpy, the combination of which determines spontaneity in biological processes, is not always feasible in introductory biology and chemistry courses.

IPLS courses can play an important role in encouraging students not just to *associate* spontaneity with a negative change in Gibbs free energy, but to understand *how* and *why* that negative sign emerges from energetic and entropic contributions. This is not to suggest that one should ignore other approaches to the Second Law, ones that consider the entropy of the surrounding universe along with that of the system. Indeed, the two approaches are of course complementary and, when employed thoughtfully, should only serve to reinforce each other. Our claim is only that we do our life science students a disservice when we do not provide them with opportunities to explicitly connect their understanding of Gibbs free energy with a formulation of the Second Law.

To begin to provide such opportunities in our NEXUS/Physics course, we designed two small group problem-solving tasks to be completed in two 50-minute class sessions over consecutive weeks. We describe some features of these activities not because we view them as final products to be adopted, but because they illustrate some ways in which task design can attend to the tensions students describe and make productive use of student resources.

The first task asked students to carefully examine why it is that oil and water do not spontaneously mix, *i.e.*, it was designed to unpack the entropic underpinnings of the hydrophobic effect. The task was scaffolded with open-ended questions prompting the students to consider differences in the molecular degrees of freedom between the state in which the oil molecules are dispersed homogenously throughout a volume of water and the state in which the oil and water regions are separate. The questions served to explicitly problematize the idea of disorder, since the state in which oil and water are homogeneously mixed exhibits disorder in the colloquial but not technical sense in this specific situation. This task gives students an explicit opportunity to grapple with the multiple possible meanings of disorder and to refine their understanding of entropy.

The second task built on the first to explore the formation of lipid bilayer cell membranes. In this task, students weighed the competing effects of energy and entropy in a qualitative way, accounting for the many factors that go into determining spontaneity for a complex, authentic biological process. In considering these energetic effects, students called upon their ideas about electrostatics in the context of a thermodynamics task, thereby linking two realms that could all too easily remain disconnected. This activity helped make explicit the tension between disorder and the formation of organized structures like the lipid bilayer cell membrane, and asked students to make use of their resources for thinking about Gibbs free energy in order to resolve the tension.

The intention of these tasks was not to arrive at a quantitative result, but rather to address tensions surrounding entropy and spontaneity, and to help students better understand the relative roles of entropy and energy in driving the separation of oil and water. Designing open-ended, discussion-provoking tasks of this nature is challenging.[27] It is inevitably a highly iterative process in which student feedback plays an essential role in task design, and in which one must be willing to be led in unexpected directions by insightful student ideas.

We have implemented the two-week task twice, in slightly different forms, and the third version will incorporate even more changes inspired by our discussions with students and by what we have learned during our two years designing an IPLS environment. We have evidence that students find these tasks to be successful in helping them build connections even in these early stages of design.[28] We therefore do not suggest that one must wait until the tasks do exactly what one intends them to do before using them productively. Rather, we provide this example as a way of encouraging IPLS instructors to consider and see as feasible the design of tasks that generate thoughtful discussions about the Second Law and Gibbs Free Energy.

## VI. CONCLUSION: TOWARD GREATER COHERENCE BETWEEN DISCIPLINARY TREATMENTS OF ENTROPY AND SPONTANEITY

The student data presented in this paper call for a treatment of entropy in IPLS courses that emphasizes its role in determining the spontaneity of processes, including biological ones. Students develop intuition for such spontaneity in their introductory biology and chemistry courses, and for the relation of spontaneity to the sign of the change in Gibbs free energy. We would be well served to leverage these intuitions when introducing the Second Law of Thermodynamics.

Unpacking the complex interplay between energy and entropy in determining the sign of the free energy change requires that we develop a set of illustrative and discussion-generating problems that help students understand these sometimes competing effects. The burden for developing such problems lies on the shoulders of *both* IPLS instructors and those teaching introductory biology and chemistry courses. The benefit of doing so is that our students will have opportunities to explore a more coherent thermodynamic world.

## ACKNOWLEDGMENTS

Many thanks to Abigail Daane and Eric Kuo for thoughtful insights about entropy and free energy, and for helping to make sense of the student ideas described in this paper. The authors also thank Chris Bauer, Melanie Cooper, Catherine Crouch, and Mike Klymkowsky for many useful interdisciplinary conversations about thermodynamics, and the UMCP Physics Education Research Group (PERG) and Biology Education Research Group (BERG) for discussions about students' reasoning about the Second Law. The slides in Figure 2 were adapted from ones produced by Julia Chan and Chris Bauer at the University of New Hampshire. This work is supported by NSF-TUES DUE 11-22818, and the HHMI NEXUS grant.